\documentclass[conference]{IEEEtran}
\usepackage{cite}
\ifCLASSINFOpdf
  \usepackage[pdftex]{graphicx}
  \graphicspath{{../figures/}}
  \DeclareGraphicsExtensions{.pdf,.jpeg,.png}
\else
\fi
\usepackage{amsmath}
\usepackage[bookmarks=false,colorlinks=true,linkcolor=blue,urlcolor=cyan]{hyperref}
\usepackage{flushend} % for last page balance

\begin{document}
%
% paper title
% Titles are generally capitalized except for words such as a, an, and, as,
% at, but, by, for, in, nor, of, on, or, the, to and up, which are usually
% not capitalized unless they are the first or last word of the title.
% Linebreaks \\ can be used within to get better formatting as desired.
% Do not put math or special symbols in the title.
\title{The Ouroboros of Memristors: Neural Networks Facilitating Memristor Programming}

% author names and affiliations
% use a multiple column layout for up to three different
% affiliations
\author{\IEEEauthorblockN{Zhenming Yu\IEEEauthorrefmark{1}\IEEEauthorrefmark{2}\IEEEauthorrefmark{3}, Ming-Jay Yang\IEEEauthorrefmark{2}\IEEEauthorrefmark{3}, Jan Finkbeiner\IEEEauthorrefmark{1}\IEEEauthorrefmark{2}, Sebastian Siegel\IEEEauthorrefmark{2}, John Paul Strachan\IEEEauthorrefmark{1}\IEEEauthorrefmark{2}, Emre Neftci\IEEEauthorrefmark{1}\IEEEauthorrefmark{2}}

\IEEEauthorblockA{\IEEEauthorrefmark{1}Fakultät für Elektrotechnik und Informationstechnik, RWTH Aachen, Aachen, 52074, Germany}
\IEEEauthorblockA{\IEEEauthorrefmark{2}Peter Grünberg Institut, Forschungszentrum Jülich GmbH, Jülich, 52425, Germany}
\IEEEauthorblockA{\{z.yu, m.yang, e.neftci\}@fz-juelich.de}
\IEEEauthorblockA{\IEEEauthorrefmark{3}These authors contributed equally to this work.}}

% \author{\IEEEauthorblockN{Anonymous Authors}

% \IEEEauthorblockA{Placeholders for}
% \IEEEauthorblockA{Blind Review}
% \IEEEauthorblockA{Anonymous Email Address}}

% conference papers do not typically use \thanks and this command
% is locked out in conference mode. If really needed, such as for
% the acknowledgment of grants, issue a \IEEEoverridecommandlockouts
% after \documentclass

% for over three affiliations, or if they all won't fit within the width
% of the page, use this alternative format:
% 
%\author{\IEEEauthorblockN{Michael Shell\IEEEauthorrefmark{1},
%Homer Simpson\IEEEauthorrefmark{2},
%James Kirk\IEEEauthorrefmark{3}, 
%Montgomery Scott\IEEEauthorrefmark{3} and
%Eldon Tyrell\IEEEauthorrefmark{4}}
%\IEEEauthorblockA{\IEEEauthorrefmark{1}School of Electrical and Computer Engineering\\
%Georgia Institute of Technology,
%Atlanta, Georgia 30332--0250\\ Email: see http://www.michaelshell.org/contact.html}
%\IEEEauthorblockA{\IEEEauthorrefmark{2}Twentieth Century Fox, Springfield, USA\\
%Email: homer@thesimpsons.com}
%\IEEEauthorblockA{\IEEEauthorrefmark{3}Starfleet Academy, San Francisco, California 96678-2391\\
%Telephone: (800) 555--1212, Fax: (888) 555--1212}
%\IEEEauthorblockA{\IEEEauthorrefmark{4}Tyrell Inc., 123 Replicant Street, Los Angeles, California 90210--4321}}

% use for special paper notices
%\IEEEspecialpapernotice{(Invited Paper)}

% make the title area
\maketitle

% As a general rule, do not put math, special symbols or citations
% in the abstract
\begin{abstract}

Memristive devices hold promise to improve the scale and efficiency of machine learning and neuromorphic hardware, thanks to their compact size, low power consumption, and the ability to perform matrix multiplications in constant time. However, on-chip training with memristor arrays still faces challenges, including device-to-device and cycle-to-cycle variations, switching non-linearity, and especially SET and RESET asymmetry ~\cite{10052010, lee2022impact}.
To combat device non-linearity and asymmetry, we propose to program memristors by harnessing neural networks that map desired conductance updates to the required pulse times. With our method, approximately $95\%$ of devices can be programmed within a relative percentage difference of $\pm50\%$ from the target conductance after just one attempt.
Our approach substantially reduces memristor programming delays compared to traditional write-and-verify methods, presenting an advantageous solution for on-chip training scenarios.
Furthermore, our proposed neural network can be accelerated by memristor arrays upon deployment, providing assistance while reducing hardware overhead compared with previous works~\cite{gokmen2020algorithm, gokmen2021enabling, onen2022neural, rasch2023fast}.

This work contributes significantly to the practical application of memristors, particularly in reducing delays in memristor programming. It also envisions the future development of memristor-based machine learning accelerators.
\end{abstract}

% \footnotetext{These authors contributed equally to this work.}

% no keywords
% keywords: 

% For peer review papers, you can put extra information on the cover
% page as needed:
% \ifCLASSOPTIONpeerreview
% \begin{center} \bfseries EDICS Category: 3-BBND \end{center}
% \fi
%
% For peerreview papers, this IEEEtran command inserts a page break and
% creates the second title. It will be ignored for other modes.
\IEEEpeerreviewmaketitle

\section{Introduction}
\label{sec:intro}
Memristive devices have emerged as promising accelerators for machine learning and neuromorphic engineering, thanks to their compact size, non-volatility, and low latency ~\cite{spiga2020memristive}. While previous studies have showcased memristor-based inference accelerators~\cite{hu2018memristor,le202364}, less progress has been seen with on-chip training and fine-tuning. While high-precision programming of memristor conductances is attainable through write-and-verify methods, as demonstrated in~\cite{rao2023thousands}, the extended programming delay hinders its application in on-chip learning and impedes the deployment of memristive crossbar arrays. In on-chip learning scenarios, pulses of specific duration are preferred~\cite{rasch2021flexible}. However, non-idealities in memristors, such as non-linearity, asymmetry, device-to-device, and cycle-to-cycle variations, with update asymmetry severely impact network performance~\cite{10052010, lee2022impact}. 
Although non-ideality-aware methods and algorithms like Tiki-Taka offer solutions to mitigate these issues, they need additional steps during training and introduce extra hardware overhead~\cite{gokmen2020algorithm, gokmen2021enabling, onen2022neural, rasch2023fast}.

\begin{figure}[!htb]
\centering
\includegraphics[width=8.3cm]{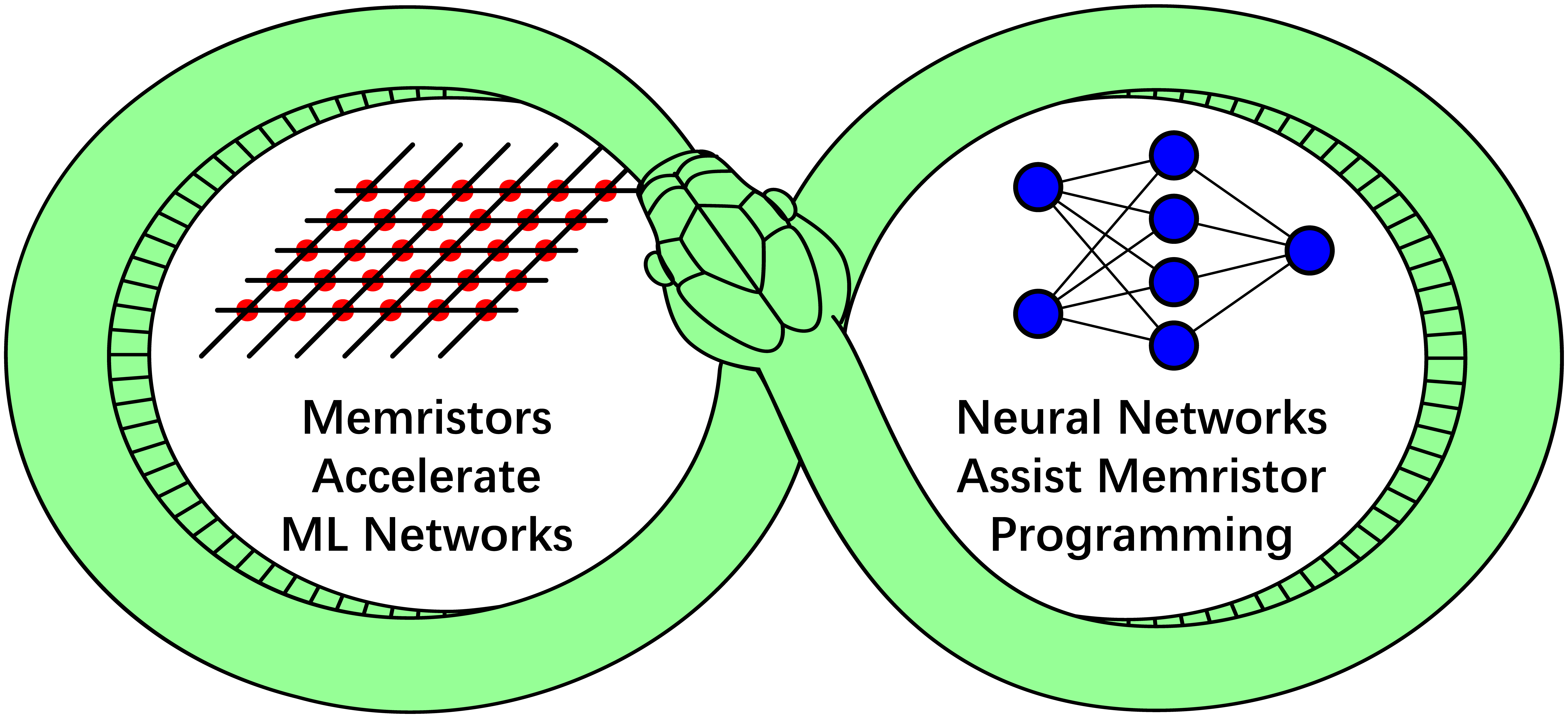}
\caption{The Ouroboros of Memristors: While memristors are typically utilized to bolster machine learning networks, our innovative approach leverages neural networks to streamline memristor programming, thereby amplifying on-chip training performance for upcoming memristor-based accelerators.}
\label{fig:Ouroboros_of_Memristors}
\end{figure}

% Beyond the conventional viewpoint of programming memristors with predefined pulses, the input-output relationship reveals a shared root cause for update asymmetry and non-linearity: attempting to program $\delta G$ while manipulating pulse time $t$ and voltage $U$. This stems from the challenge of memristor programming. 
Upon revisiting the method of programming memristors using predefined pulses, an intriguing observation emerges: in our efforts to program $\delta G$, we manipulate pulse time $t$ and voltage $U$, which can only impact $G$ through complex physical dynamics. Naturally, the resulting response is neither linear nor symmetric. To effectively address the challenges posed by update asymmetry and non-linearity, what we really need is something capable of translating the desired $\delta G$ into the required pulse time $t$.
However, due to inherent noise and variations in memristors, constructing static look-up tables is challenging. To address this obstacle, we propose a novel method: training neural networks for memristor programming to map desired conductance updates to the required voltage pulses. As the switching dynamics of memristors are linked to their current conductance, the neural pulse predictor takes the current conductance $G$ and the update $\Delta G$ as input. This symbiotic relationship, depicted in Fig.~\ref{fig:Ouroboros_of_Memristors}, where memristors accelerate the neural pulse predictor, which in turn aids in programming themselves, holds significant potential. 
Given that the programming pulse prediction network consistently operates in inference mode, it can be embedded onto memristor arrays through precise write-and-verify methods, producing swift prediction results while utilizing only a fraction of the hardware resources.
% In Sec.~\ref{sec:simulation_setup}, we introduce the simulation environment, followed by details about our neural pulse predictor in Sec.~\ref{sec:neural_network}. In Sec.~\ref{sec:demo}, we demonstrate the performance of our networks through a simple write-and-verify demo.
%
% no \IEEEPARstart
%
\section{Simulation Setup}
\label{sec:simulation_setup}

\subsection{Device Modeling}
\label{sec:model}
\begin{figure}[!htb]
\centering
\includegraphics[width=\hsize]{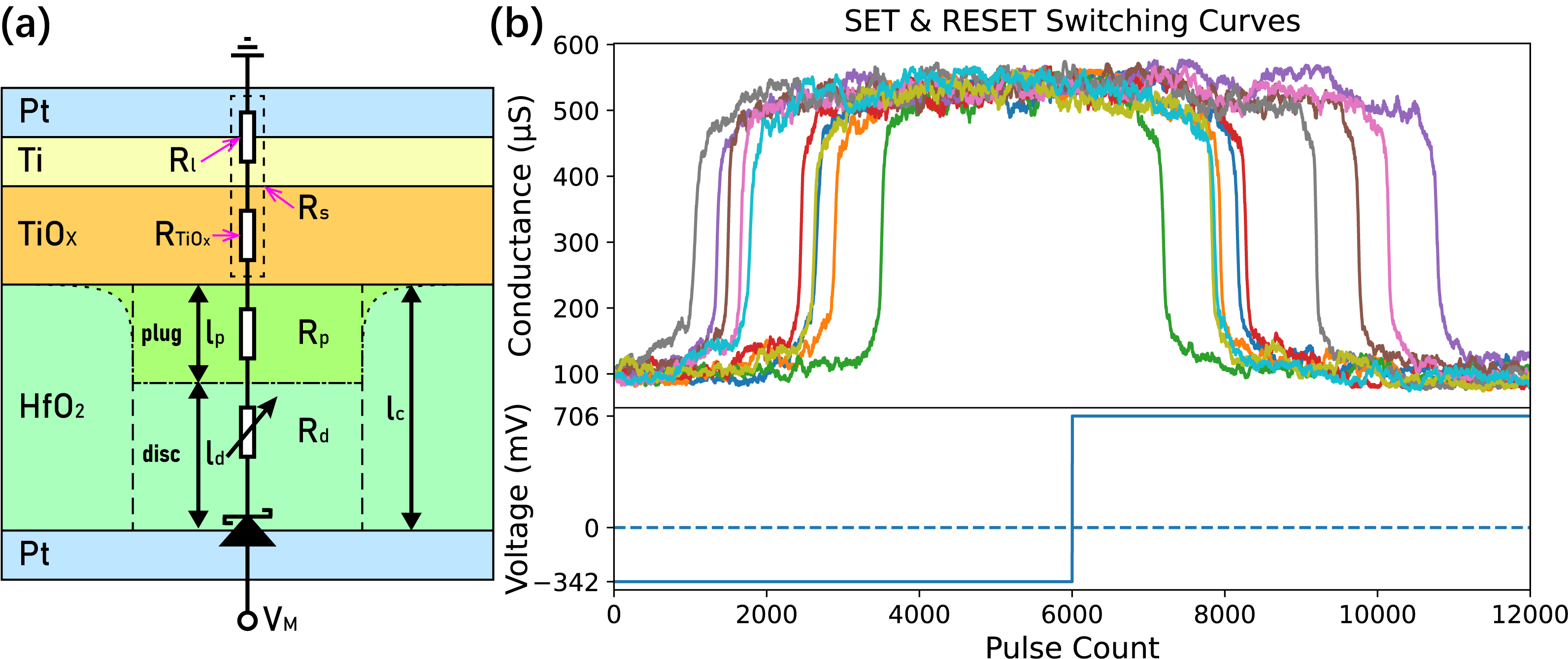}
\caption{\textbf{(a)}: Illustration of the equivalent circuit diagram for the JART memristor model. \textbf{(b)}: Simulation results depicting switching curves following the application of 6000 SET pulses and 6000 RESET pulses across 10 devices.}
\label{fig:JART_model}
\end{figure}

In accordance with~\cite{10052010}, we employed the JART VCM model~\cite{cuppers2019exploiting} for our simulation setup. This model was chosen for its detailed physics-driven noise mechanisms and has been experimentally validated~\cite{bengel2020variability}. As illustrated in Fig.~\ref{fig:JART_model}(a), the JART model abstracts the filament region of VCM devices as a stack, represented as circuit elements in series. Cycle-to-cycle noises were implemented as random walks in the length and radius of the filament region, as well as in the boundaries of oxygen vacancy concentration, which induces noises in the $G_\text{max}$ and $G_\text{min}$. We adopted a simplified model ~\cite{10192107} mathematically fitted to the JART model to streamline calculations. The simulation results are presented in Fig.~\ref{fig:JART_model}(b).
% Note that our pulse predictor network is independent to the device model. Here the JART model is chosen to generate dataset for training and validating the predictor.

\subsection{Dataset Generation}
\label{sec:data}
We utilized the model detailed in Sec.~\ref{sec:model} to generate the training dataset. We selected only one pair of $V_\text{SET}$ and $V_\text{RESET}$ for programming the memristor. These voltages were maintained following the specifications in ~\cite{10052010}, ensuring smooth transitions and accessibility to intermediate states.

The dataset generation process begins with a random conductance $G_\text{start}$ and a random target $G_\text{target}$ within the operational range. We then choose to apply $V_\text{SET}$ or $V_\text{RESET}$ based on the direction of the target and apply the voltage for $10ns$. Cycle-to-cycle noise was introduced to the selected parameters, as explained in Sec.~\ref{sec:model}, and the conductance was recorded. The cycle continues until the conductance before and after the time step lies on two sides of $G_\text{target}$. The conductance closer to $G_\text{target}$ was selected as $G_\text{end}$, and the time for the device to switch from $G_\text{start}$ to $G_\text{end}$ was recorded as $t_\text{pulse}$.

It is important to note that in our simulation setup, we did not account for the conductance change during the rising and falling edges. Thus, many short pulses are equivalent to one long pulse. As we could check the device conductance at any time in simulation without disturbing the memristor state variables, we effectively have a prolonged pulse that stopped just when $G_\text{end}$ was closest to $G_\text{target}$.

Given that we only used 2 applied voltages, $V_\text{SET}$ and $V_\text{RESET}$, we employ the sign of $t_\text{pulse}$ to denote the switching direction. If $G_\text{target}$ exceeded $G_\text{start}$, $V_\text{SET}$ was used, and $t_\text{pulse}$ remained positive. Conversely, if $G_\text{target}$ was less than $G_\text{start}$, $V_\text{RESET}$ was used, and $t_\text{pulse}$ was designated as negative.

We recorded the complete $G-t$ histories for a duration of $2 \times t_\text{pulse}$ as references during both training and validation. This process was repeated 10,000 times for the complete dataset, which was then partitioned into ratios of $[80\%, 10\%, 10\%]$ for training, validation, and testing, respectively.

\section{Neural Pulse Predictor}
\label{sec:neural_network}
\subsection{Network Training}
\label{sec:training}
We used PyTorch to construct and train our neural pulse predictor. The inputs, $G_\text{start}$ and $\delta G$, as well as the output, $t_\text{pulse}$, were approximately normalized to the interval [0, 1]. Due to the cycle-to-cycle noises, the input range was not precisely defined. Therefore, we opted for a rough normalization, bringing the values to similar ranges. We designed a fully connected neural network with 3 hidden layers, configured with sizes [2 (input size), 32, 64, 32, 1 (output size)].
The Relative Percentage Difference (RPD) defined as
\begin{equation}
\label{eq:RPD}
    \text{RPD}=\text{mean} \left(\frac{\left|X_\text{output}-X_\text{target}\right|}{X_\text{target}} \right)
\end{equation}
was used to assess the performance. The network was trained for 100 epochs using Mean Squared Error (MSE) loss between $t_\text{output}$ and $t_\text{target}$. We saved networks based on their RPD on the validation dataset. Two approaches for RPD evaluation were explored: direct evaluation with $t$ and mapping onto $G$. In the first approach, RPD was calculated with $t_\text{output}$ and $t_\text{target}$. In the second approach, $t_\text{output}$ was mapped to $G_\text{output}$ using the $G-t$ histories recorded. If $t_\text{output}$ fell outside the recorded range, the first or last recorded $G$ was used as $G_\text{output}$. Subsequently, RPD was calculated using $G_\text{output}$ and $G_\text{target}$.

To evaluate the network's performance, we deployed it in simulation as described in Sec.~\ref{sec:simulation_setup}. The network received the current conductance $G_\text{start}$ and the target $\delta G$, predicting the voltage pulse needed to program $G_\text{target}$ onto the memristor device. We simulated the pulse accordingly, compared $G_\text{end}$ with $G_\text{target}$, and calculated RPD using Eq.~\ref{eq:RPD}. Additionally, we included a baseline performance calculated by utilizing the pulse time that would result in the same $\delta G$ with the noise-free model unaffected by cycle-to-cycle noises.

\begin{figure}[!htb]
\centering
\includegraphics[width=\hsize]{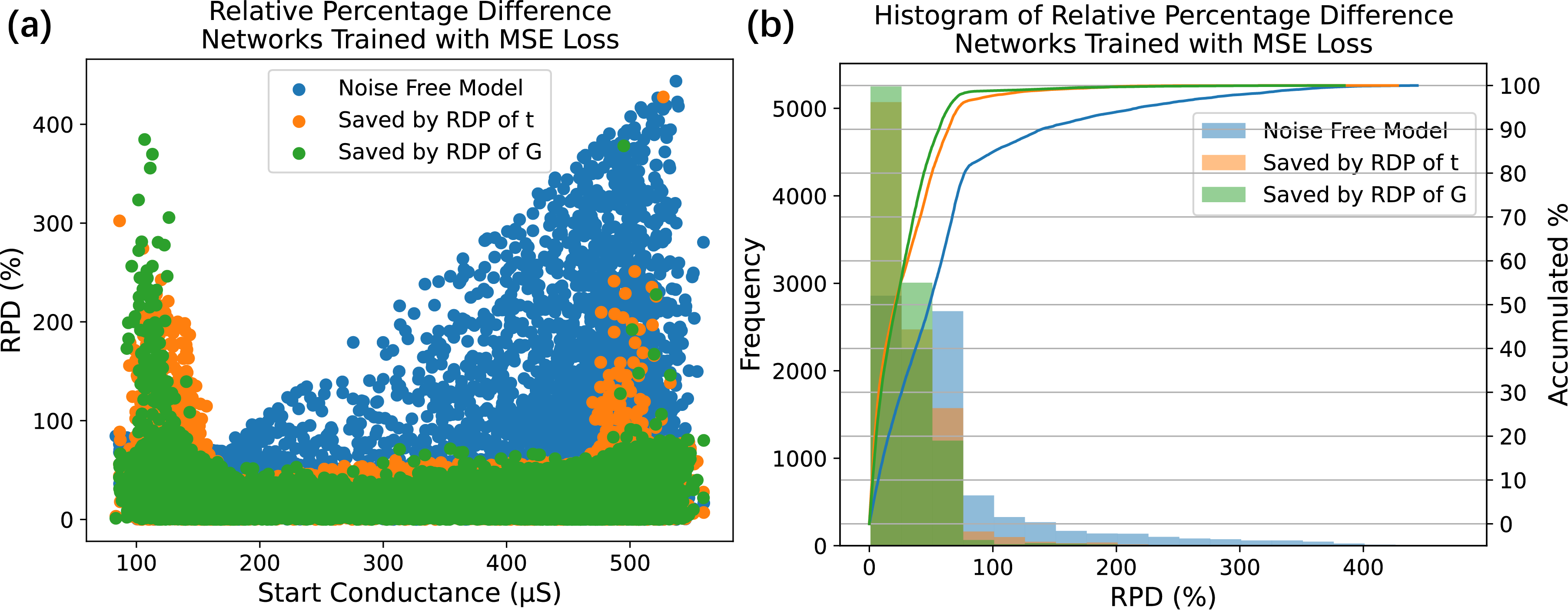}
\caption{\textbf{(a)} RPDs of $G_\text{output}$ generated by networks deployed in simulation and \textbf{(b)} the corresponding histograms. These networks underwent training from the ground up using MSE loss. The models were saved based on the best RPD of $t_\text{output}$ or $G_\text{output}$ observed on the validation dataset.}
\label{fig:RPD_his_from_scratch}
\end{figure}

The networks trained with MSE loss and saved based on the best RPD relative to $t_\text{target}$ or $G_\text{target}$ on the validation dataset exhibited similar performance. Both networks outperformed the baseline pulse predictor running a noise-free model, as shown in Fig.~\ref{fig:RPD_his_from_scratch}. Fig.~\ref{fig:RPD_his_from_scratch}(a) reveals a strong association between the performance of the trained models and the starting conductance. When $G_\text{start}$ is in the middle of the full conductance range, the trained networks perform well, programming $G_\text{end}$s closer to $G_\text{target}$ than the noise-free baseline. However, when $G_\text{start}$ is close to $G_\text{min}$ or $G_\text{max}$, errors in conductance programming begin to increase.

This behavior is attributed to the memristor model described in Sec.~\ref{sec:model}, which features two self-accelerating feedback loops in its switching dynamics. In the SET direction, the rise of conductance induces more thermal heat-up, lowering the energy barrier for device switching. In the RESET direction, the rise of resistance increases the voltage drop on the memristor device, accelerating the switching process. Both self-accelerating processes saturate when the device approaches the conductance boundaries, resulting in S-shaped switching curves shown in Fig.~\ref{fig:JART_model}(b). As a result, the model reacts slowly to pulses when it is close to $G_\text{min}$ and $G_\text{max}$. With the noise magnitude being bigger than the conductance changes, shorter pulses may not appear to have significant impacts on the recorded data. Consequently, the neural pulse predictor tends to produce larger errors when starting with conductance close to $G_\text{min}$ or $G_\text{max}$.
\begin{figure}[!htb]
\centering
\includegraphics[width=\hsize]{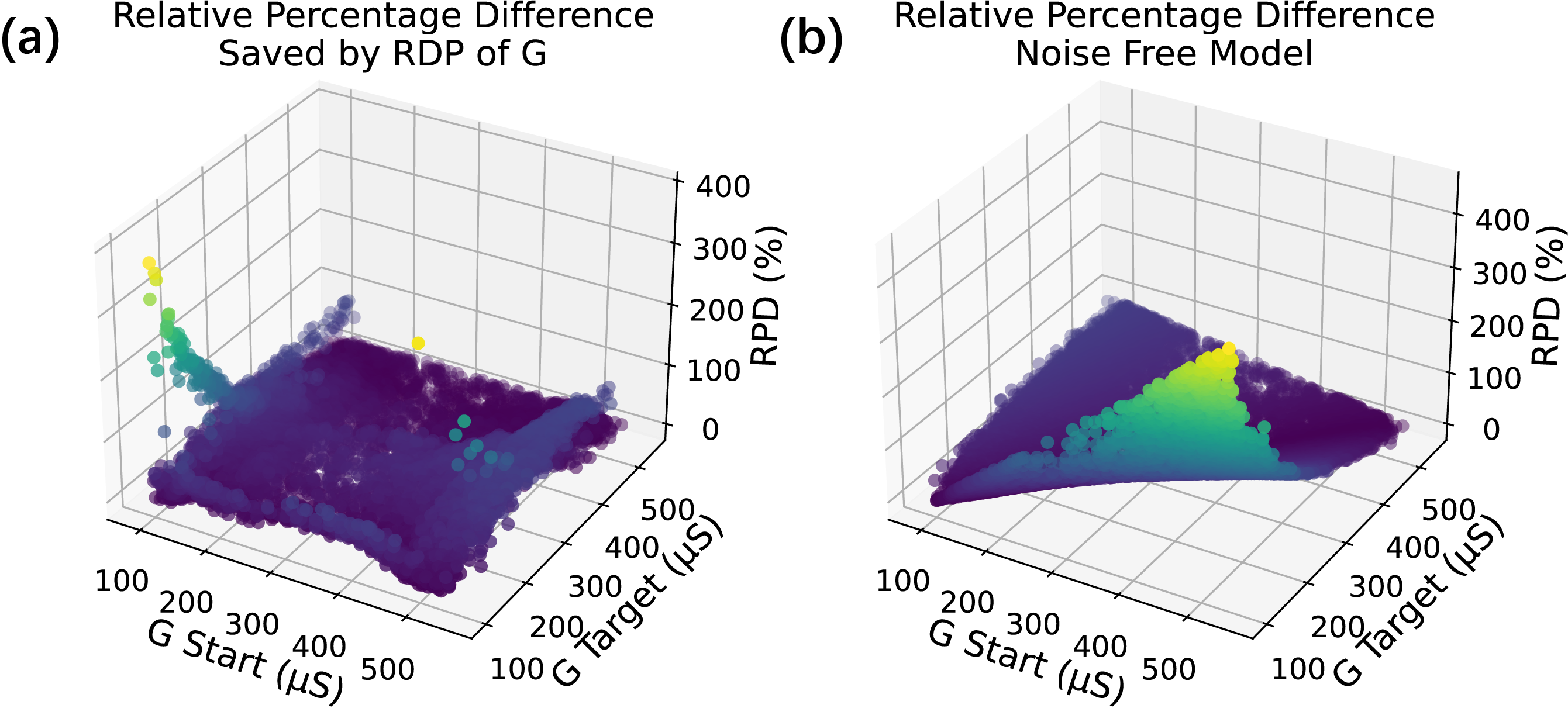}
\caption{RPDs plotted with $G_\text{start}$ and $G_\text{target}$ for the pulse predictors \textbf{(a)} trained from scratch with MSE loss, and \textbf{(b)} based on noise-free model.}
\label{fig:3D_his_from_scratch}
\end{figure}

A closer examination of the results, plotted with $G_\text{target}$ in Fig.~\ref{fig:3D_his_from_scratch}, reveals an increase in RPD at low $G_\text{start}$ with the trained neural pulse predictor, but not with the noise-free baseline. This discrepancy arises because we encoded pulse directions in the sign of $t_\text{pulse}$. In regions where $G_\text{start}$ is close to $G_\text{min}$, the effect of pulses in both directions is insignificant, and noise dominates the conductance change. In such cases, the trained network has a higher likelihood of predicting pulses in the wrong direction. However, the noise-free baseline is calculated using simulation and never has direction error. This provides the noise-free baseline with an advantage, allowing it to outperform the trained networks with low $G_\text{start}$.

\subsection{Network Fine-tuning}
\label{sec:fine-tuning}
If we look at the training method in Sec.~\ref{sec:training}, an intriguing aspect becomes apparent. While the goal is to bring $G_\text{end}$ closer to $G_\text{target}$, the networks are trained on $t_\text{target}$. While this method is both straightforward and computationally simple, it can not guarantee proximity between $G_\text{end}$ and $G_\text{target}$ due to the model's highly non-linear $G-t$ transfer curve as evident from Fig.~\ref{fig:JART_model}(b).

\begin{figure}[!htb]
\centering
\includegraphics[width=\hsize]{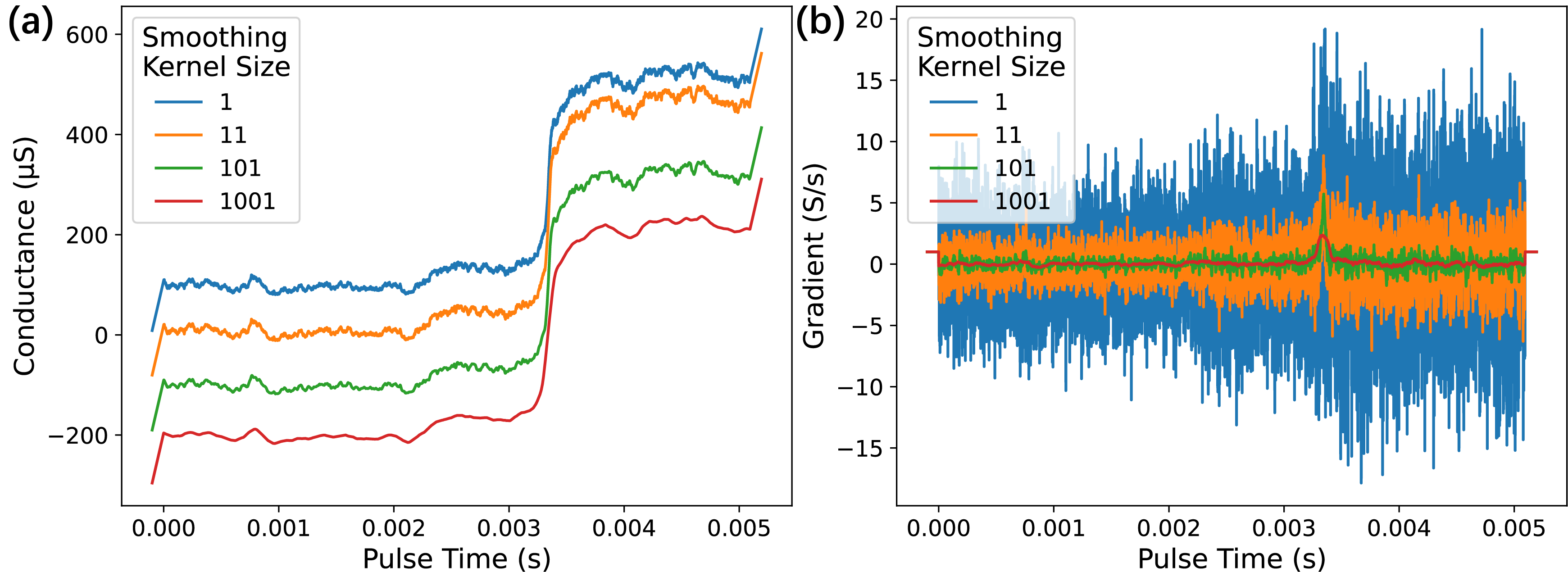}
\caption{\textbf{(a)} Mapping of $G-t$ (kernel size 11, 101, and 1001 shifted for better visibility) and \textbf{(b)} the corresponding gradients in an example $G-t$ history. Various kernel sizes were employed during the smoothing process.}
\label{fig:g_grad}
\end{figure}

In light of this, an alternative approach was explored, utilizing the recorded $G-t$ histories as piecewise functions to map $t_\text{output}$ to $G_\text{output}$. The network was then trained on MSE loss between $G_\text{output}$ and $G_\text{target}$. To mitigate noise in the recorded $G-t$ histories, we implemented a smoothing process using moving average filters with various kernel sizes for convolution. Customized backward function was used to mitigate non-differentiable points, and clamped the gradient to 1 to guide the output towards the recorded regions, as depicted in Fig.~\ref{fig:g_grad}.

\begin{figure}[!htb]
\centering
\includegraphics[width=\hsize]{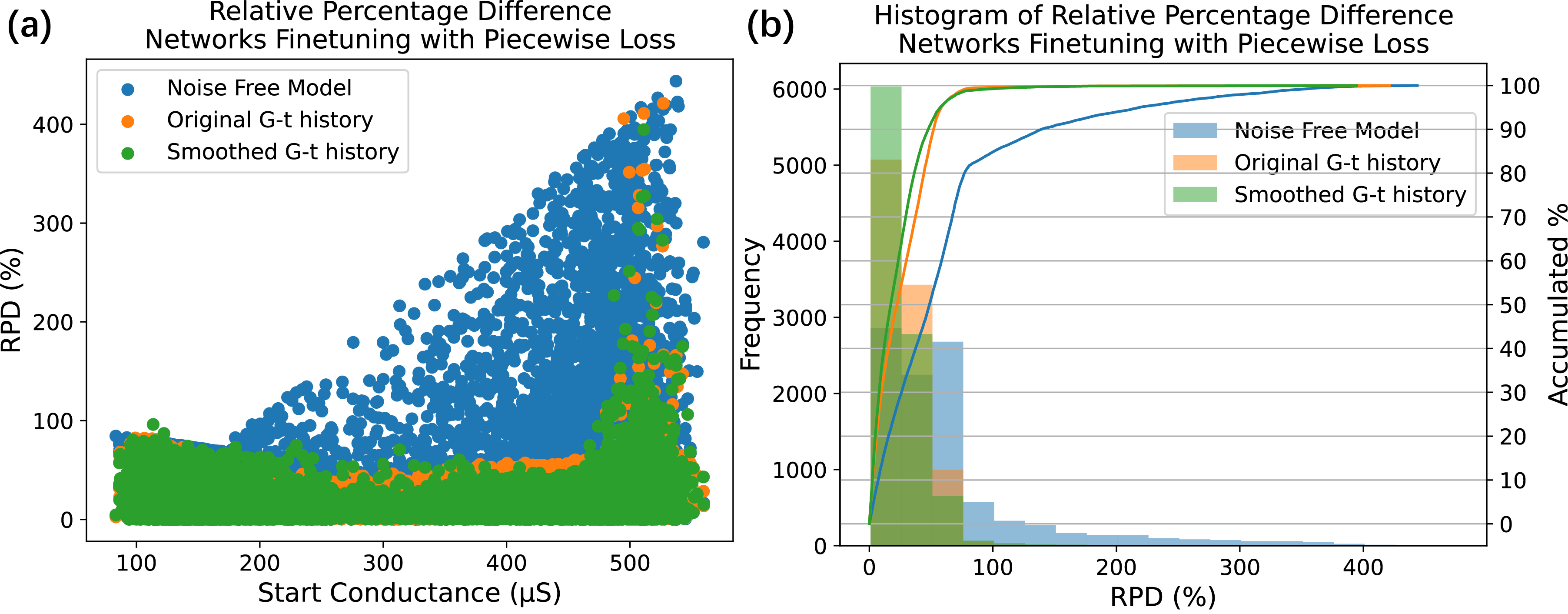}
\caption{\textbf{(a)} RPDs of $G_\text{output}$ generated by various networks deployed in simulation and \textbf{(b)} the corresponding histograms depicting the distribution. These networks underwent fine-tuning from a model initially trained with MSE loss and were saved based on the best RPD performance observed on the validation dataset.}
\label{fig:RPD_his_fine_tuning}
\end{figure}

We utilized the model from Fig.~\ref{fig:3D_his_from_scratch}(a) and fine-tuned the weights. We first attempted fine-tuning without smoothing the $G-t$ histories, training the network for 100 epochs and saving the weights that produced the best match between $G_\text{output}$ and $G_\text{target}$ on the validation dataset for testing. Then, we experimented with smoothed $G-t$ histories by decreasing smoothing kernel sizes from 1001, 101, 11 to 1 during the training process. The network was trained with each smoothing kernel for 50 epochs, and the weights were saved based on the best RPD performance during validation. This approach aimed to address overshooting gradient descents, stabilize the training process, and enhance the overall performance, and the results are shown in Fig.~\ref{fig:RPD_his_fine_tuning}.

The fine-tuning process yielded a notable improvement in performance. By utilizing fixed gradients outside the recorded range, the output was guided toward the correct direction, resulting in performance increases at low $G_\text{start}$, matching the noise-free baseline, as shown in Fig.~\ref{fig:RPD_his_fine_tuning}(a). Performance was further enhanced with smoothed gradients, and RPDs of less than $50\%$ were achieved in approximately $95\%$ of all trials.

\section{Write-and-Verify Demo}
\label{sec:demo}
\subsection{Demonstration In Simulation}
\label{sec:simulation}
\begin{figure}[!htb]
\centering
\includegraphics[width=\hsize]{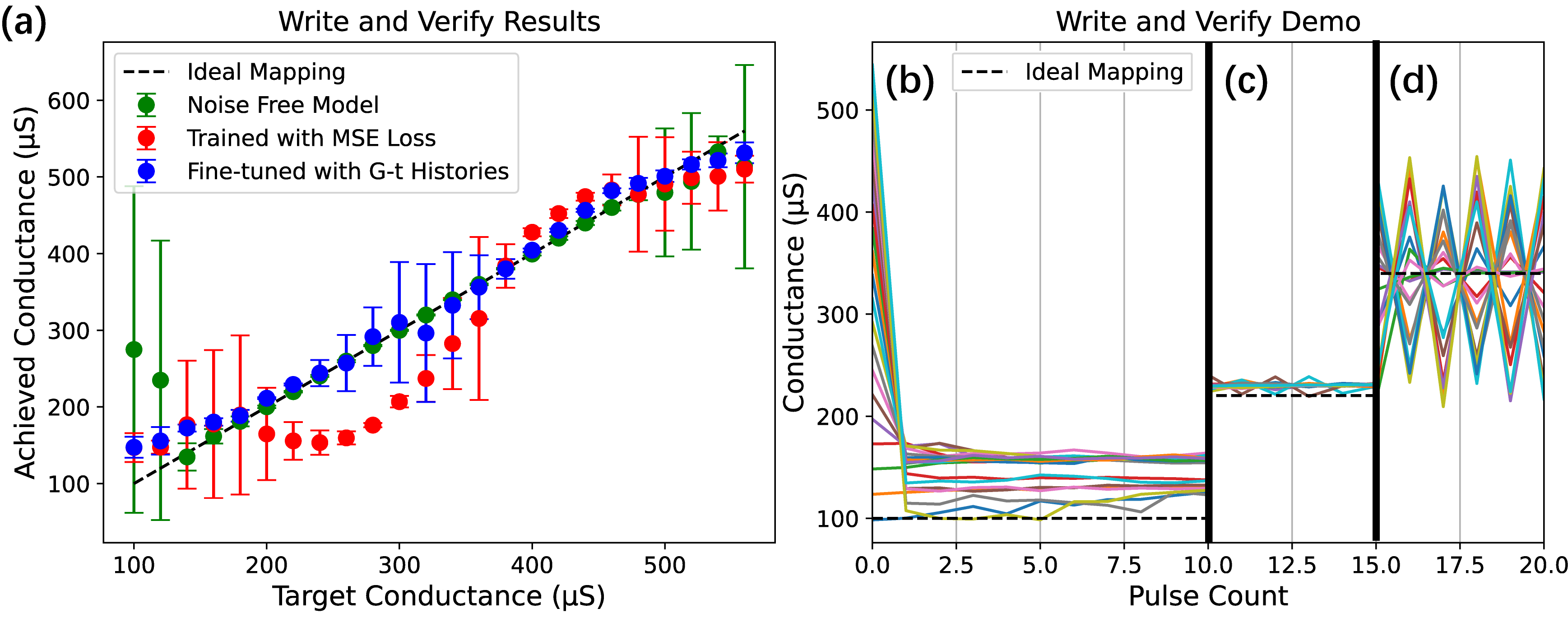}
\caption{\textbf{(a)} Converged mapping of $G_\text{end}$ and $G_\text{target}$ after 20 write-and-verify iterations. \textbf{(b)} Histories of the predictor network fine-tuned with smoothed $G-t$ with $G_\text{target}$ of \textbf{(b)} $100\mu S$, \textbf{(c)} $220\mu S$, and \textbf{(d)} $340\mu S$.}
\label{fig:W_and_V_results}
\end{figure}

To showcase the performance of our networks, we configured the memristor device to different conductances across the range and attempted to program them towards a shared $G_\text{target}$. After 10 write-and-verify iterations, most devices reached their converged conductance. The results after 20 iterations are presented in Fig.~\ref{fig:W_and_V_results}(a). 
The noise-free model faces challenges near $G_{\text{min}}$ and $G_{\text{max}}$ due to the self-accelerating processes discussed in Sec.~\ref{sec:training}, resulting in its poor performance. Although it successfully converges in other regions, the simulation time is significantly extended due to intensive computations. In contrast, the model fine-tuned with piecewise $G-t$ histories achieves performance comparable to the baseline. It achieved stability with low standard deviations and aligns well with $G_{\text{target}}$. The model trained with MSE loss on $t_{\text{target}}$ exhibits noisier behavior and lacks consistent convergence to $G_{\text{target}}$.

For $G_\text{target}$ close to the conductance boundaries, $G_\text{end}$ is heavily influenced by the noise of $G_\text{min}$ and $G_\text{max}$, as shown in Fig.~\ref{fig:W_and_V_results}(b), $G_\text{end}$ stabilize slightly away from $G_\text{target}$. In regions where the network performs well, as Fig.~\ref{fig:W_and_V_results}(c) shows, $G_\text{end}$ stay stably close to $G_\text{target}$. However, towards where the $t_G$ transfer curve becomes sharp and the impact of noise increases, network performance starts to decline, with $G_\text{end}$ overshooting and oscillating around $G_\text{target}$, as shown in Fig.~\ref{fig:W_and_V_results}(d).

\begin{figure}[!htb]
\centering
\includegraphics[width=\hsize]{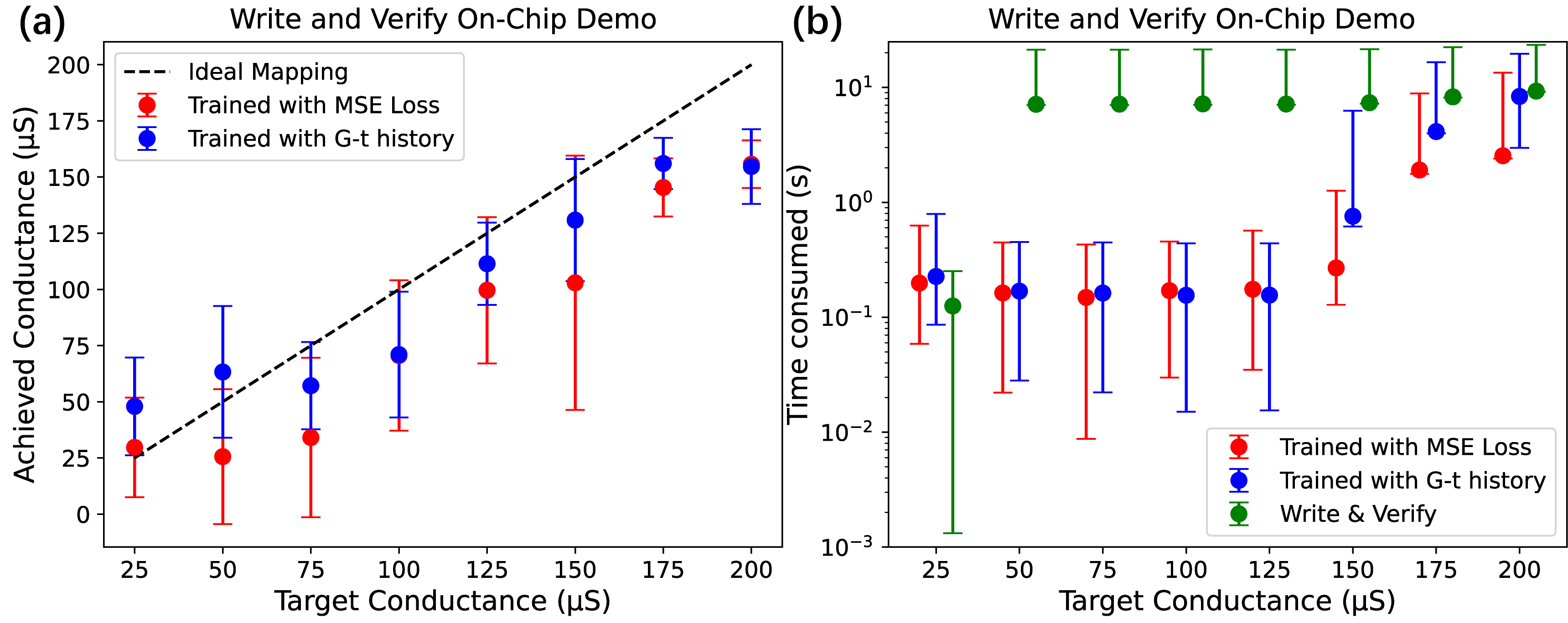}
\caption{On-chip demonstration of \textbf{(a)} converged mapping of $G_\text{end}$ and $G_\text{target}$ after 10 iterations. \textbf{(b)} comparison of programming delay for the device to converge into the window of $G_\text{target} \pm 50\mu S$ using different write methods.}
\label{fig:On_chip_W_and_V_results}
\end{figure}

\subsection{On-Chip Demonstration}
\label{sec:on_chip}
To demonstrate the predictor's performance on real devices, we utilize a memristor testing system integrated with CMOS circuitry \cite{superT}. The device characteristics were fitted with a statistical model \cite{ICONS}, and the neural predictor was trained as outlined in Section \ref{sec:neural_network}. The obtained results are presented in Fig. \ref{fig:On_chip_W_and_V_results}. %Due to the Sim2Real transfer problem, we encounter a performance gap when compared with results in simulation. 
Notably, the time delay for writing target conductance is significantly lower than the baseline write-and-verify method \cite{IPbaseline}.
Precision performance is expected to further improve with data from models that better account for %cycle-to-cycle and device-to-device 
device and circuit noises, or directly from experimental measurements.

\section{Discussion}
\label{sec:discussion}
\textbf{Poor Performance of Noise-free Model:}
In Sec.~\ref{sec:training}, the noise-free baseline exhibits poor performance. The exact reason for this remains unclear, but could be linked to the switching dynamics discussed in Sec.~\ref{sec:training}, where injected noises could further assist the self-accelerating switching processes, and introduce biases in the predictions.

Another plausible factor is the asymmetric impact of noise. As explained in Sec.~\ref{sec:model}, the change parameters could disproportionately influence switching dynamics, causing deviations with the predictions derived from mean values.

% \textbf{Training from Scratch With Smoothed $G-t$ Histories:}
% While it is feasible to train networks directly from scratch using smoothed $G-t$ histories (as described in Sec.~\ref{sec:fine-tuning}), this approach tends to exhibit instability during training. The gradient descent process frequently overshoots, resulting in a rise in training loss and RPD on the validation dataset, particularly when decreasing the smoothing kernel size.

% \textbf{Training with Measured Device Data:}
% While using measured device data for training is possible, it poses challenges in terms of acquiring sufficient data for convergence without overfitting. In practice, exploration of models fitted to device data would be a more viable option.

\textbf{Training with Varying Voltages:}
We trained the network with fixed voltages to ease the data collection. While high voltages may damage the device, prolonged pulse length can be achieved easily. In reality, precision tuning of applied voltages using a DAC is more feasible, whereas the precision of pulse length is limited to clock frequencies, and one could train networks that work with varying voltages instead.

\textbf{Better Write-and-Verify Approaches:}
Our approach minimizes programming overhead compared to traditional algorithms. Its simplicity offers advantages on device tuning, % in terms of latency
in terms of latency and power dissipation, beneficial for on-chip training. Although this approach might have limited precision, as long as the update sign remains unchanged, the weight will eventually converge.

\section{Conclusion}
We have introduced a novel approach to program memristor by training neural networks predicting update pulses. Our neural pulse predictor demonstrates a significant reduction in programming delay compared to traditional write-and-verify methods,
 % Our neural pulse predictor demonstrates adaptive tuning in an online fashion compared to traditional write-and-verify methods, 
 particularly advantageous in applications such as on-chip training and fine-tuning. Upon deployment, the neural pulse predictor can be integrated into memristor accelerators, utilizing a minimal fraction of the available memristor arrays while predicting pulses with an O(1) time complexity. Additionally, multiple networks can be trained to operate in parallel and enhance precision across various conductance ranges. Our work offers a fresh perspective on the symbiotic relationship between memristors and neural networks and sets the stage for innovation in memristor tuning optimizations.

% conference papers do not normally have an appendix

% use section* for acknowledgment
%  \section*{Acknowledgment}
% Place Holder for Acknowledgment Section.

% /

% /

% /

% Used only during the blind review.
 \section*{Acknowledgment}
This work was sponsored by the Federal Ministry of Education, Germany (project NEUROTEC-II grant no. 16ME0398K and 16ME0399). We thank Dr. Vasileios Ntinas for his help with the simplified model~\cite{10192107} and Dr. Stephan Menzel for his advice on the JART model noise implementations.

 \section*{Copyright}
© 2024 IEEE. Personal use of this material is permitted. Permission from IEEE must be obtained for all other uses, in any current or future media, including reprinting/republishing this material for advertising or promotional purposes, creating new collective works, for resale or redistribution to servers or lists, or reuse of any copyrighted component of this work in other works.
% \newpage

% trigger a \newpage just before the given reference
% number - used to balance the columns on the last page
% adjust value as needed - may need to be readjusted if
% the document is modified later
%\IEEEtriggeratref{8}
% The "triggered" command can be changed if desired:
%\IEEEtriggercmd{\enlargethispage{-5in}}

% references section
% can use a bibliography generated by BibTeX as a .bbl file
% BibTeX documentation can be easily obtained at:
% http://mirror.ctan.org/biblio/bibtex/contrib/doc/
% The IEEEtran BibTeX style support page is at:
% http://www.michaelshell.org/tex/ieeetran/bibtex/
%\bibliographystyle{IEEEtran}
% argument is your BibTeX string definitions and bibliography database(s)
\bibliographystyle{IEEEtran}
\bibliography{IEEEtran/citations.bib}

% Generated by IEEEtran.bst, version: 1.14 (2015/08/26)
\begin{thebibliography}{10}
\providecommand{\url}[1]{#1}
\csname url@samestyle\endcsname
\providecommand{\newblock}{\relax}
\providecommand{\bibinfo}[2]{#2}
\providecommand{\BIBentrySTDinterwordspacing}{\spaceskip=0pt\relax}
\providecommand{\BIBentryALTinterwordstretchfactor}{4}
\providecommand{\BIBentryALTinterwordspacing}{\spaceskip=\fontdimen2\font plus
\BIBentryALTinterwordstretchfactor\fontdimen3\font minus \fontdimen4\font\relax}
\providecommand{\BIBforeignlanguage}[2]{{%
\expandafter\ifx\csname l@#1\endcsname\relax
\typeout{** WARNING: IEEEtran.bst: No hyphenation pattern has been}%
\typeout{** loaded for the language `#1'. Using the pattern for}%
\typeout{** the default language instead.}%
\else
\language=\csname l@#1\endcsname
\fi
#2}}
\providecommand{\BIBdecl}{\relax}
\BIBdecl

\bibitem{10052010}
Z.~Yu, S.~Menzel, J.~P. Strachan, and E.~Neftci, ``Integration of physics-derived memristor models with machine learning frameworks,'' in \emph{2022 56th Asilomar Conference on Signals, Systems, and Computers}, 2022, pp. 1142--1146.

\bibitem{lee2022impact}
C.~Lee, K.~Noh, W.~Ji, T.~Gokmen, and S.~Kim, ``Impact of asymmetric weight update on neural network training with tiki-taka algorithm,'' \emph{Frontiers in neuroscience}, vol.~15, p. 767953, 2022.

\bibitem{gokmen2020algorithm}
T.~Gokmen and W.~Haensch, ``Algorithm for training neural networks on resistive device arrays,'' \emph{Frontiers in neuroscience}, vol.~14, p. 103, 2020.

\bibitem{gokmen2021enabling}
T.~Gokmen, ``Enabling training of neural networks on noisy hardware,'' \emph{Frontiers in Artificial Intelligence}, vol.~4, p. 699148, 2021.

\bibitem{onen2022neural}
M.~Onen, T.~Gokmen, T.~K. Todorov, T.~Nowicki, J.~A. Del~Alamo, J.~Rozen, W.~Haensch, and S.~Kim, ``Neural network training with asymmetric crosspoint elements,'' \emph{Frontiers in Artificial Intelligence}, vol.~5, p. 891624, 2022.

\bibitem{rasch2023fast}
M.~J. Rasch, F.~Carta, O.~Fagbohungbe, and T.~Gokmen, ``Fast offset corrected in-memory training,'' \emph{arXiv preprint arXiv:2303.04721}, 2023.

\bibitem{spiga2020memristive}
S.~Spiga, A.~Sebastian, D.~Querlioz, and B.~Rajendran, \emph{Memristive Devices for Brain-Inspired Computing: From Materials, Devices, and Circuits to Applications-Computational Memory, Deep Learning, and Spiking Neural Networks}.\hskip 1em plus 0.5em minus 0.4em\relax Woodhead Publishing, 2020.

\bibitem{hu2018memristor}
M.~Hu, C.~E. Graves, C.~Li, Y.~Li, N.~Ge, E.~Montgomery, N.~Davila, H.~Jiang, R.~S. Williams, J.~J. Yang \emph{et~al.}, ``Memristor-based analog computation and neural network classification with a dot product engine,'' \emph{Advanced Materials}, vol.~30, no.~9, p. 1705914, 2018.

\bibitem{le202364}
M.~Le~Gallo, R.~Khaddam-Aljameh, M.~Stanisavljevic, A.~Vasilopoulos, B.~Kersting, M.~Dazzi, G.~Karunaratne, M.~Br{\"a}ndli, A.~Singh, S.~M. Mueller \emph{et~al.}, ``A 64-core mixed-signal in-memory compute chip based on phase-change memory for deep neural network inference,'' \emph{Nature Electronics}, vol.~6, no.~9, pp. 680--693, 2023.

\bibitem{rao2023thousands}
M.~Rao, H.~Tang, J.~Wu, W.~Song, M.~Zhang, W.~Yin, Y.~Zhuo, F.~Kiani, B.~Chen, X.~Jiang \emph{et~al.}, ``Thousands of conductance levels in memristors integrated on cmos,'' \emph{Nature}, vol. 615, no. 7954, pp. 823--829, 2023.

\bibitem{rasch2021flexible}
M.~J. Rasch, D.~Moreda, T.~Gokmen, M.~Le~Gallo, F.~Carta, C.~Goldberg, K.~El~Maghraoui, A.~Sebastian, and V.~Narayanan, ``A flexible and fast pytorch toolkit for simulating training and inference on analog crossbar arrays,'' in \emph{2021 IEEE 3rd international conference on artificial intelligence circuits and systems (AICAS)}.\hskip 1em plus 0.5em minus 0.4em\relax IEEE, 2021, pp. 1--4.

\bibitem{cuppers2019exploiting}
F.~C{\"u}ppers, S.~Menzel, C.~Bengel, A.~Hardtdegen, M.~Von~Witzleben, U.~B{\"o}ttger, R.~Waser, and S.~Hoffmann-Eifert, ``Exploiting the switching dynamics of hfo2-based reram devices for reliable analog memristive behavior,'' \emph{APL materials}, vol.~7, no.~9, p. 091105, 2019.

\bibitem{bengel2020variability}
C.~Bengel, A.~Siemon, F.~C{\"u}ppers, S.~Hoffmann-Eifert, A.~Hardtdegen, M.~von Witzleben, L.~Hellmich, R.~Waser, and S.~Menzel, ``Variability-aware modeling of filamentary oxide-based bipolar resistive switching cells using spice level compact models,'' \emph{IEEE Transactions on Circuits and Systems I: Regular Papers}, vol.~67, no.~12, pp. 4618--4630, 2020.

\bibitem{10192107}
V.~Ntinas, D.~Patel, Y.~Wang, I.~Messaris, V.~Rana, S.~Menzel, A.~Ascoli, and R.~Tetzlaff, ``A simplified variability-aware vcm memristor model for efficient circuit simulation,'' in \emph{2023 19th International Conference on Synthesis, Modeling, Analysis and Simulation Methods and Applications to Circuit Design (SMACD)}, 2023, pp. 1--4.

\bibitem{superT}
C.~Li, J.~Ignowski, X.~Sheng, R.~Wessel, B.~Jaffe, J.~Ingemi, C.~Graves, and J.~P. Strachan, ``Cmos-integrated nanoscale memristive crossbars for cnn and optimization acceleration,'' \emph{IEEE International Memory Workshop}, 2020.

\bibitem{ICONS}
M.-J. Yang and J.~P. Strachan, ``State-space modeling and tuning of memristors for neuromorphic computing applications,'' \emph{ICONS '23: Proceedings of the 2023 International Conference on Neuromorphic Systems}, 2023.

\bibitem{IPbaseline}
S.~Yu, ``Neuro-inspired computing with emerging non-volatile memory,'' \emph{Proceedings of the IEEE}, 2018.

\end{thebibliography}

\end{document}